\@twosidetrue
\@twocolumntrue
\@mparswitchtrue
\def\ds@draft{\overfullrule 5pt}
\def\ds@twocolumn{\@twocolumntrue}
\def\ds@onecolumn{\@twocolumnfalse}
\newif\ifSFB@landscape
\def\ds@landscape{\SFB@landscapetrue}
\newif\ifSFB@galley
\def\ds@galley{\SFB@galleytrue}
\newif\ifSFB@referee
\def\ds@referee{%
 \SFB@refereetrue
 \@twocolumnfalse
}
\@options
\ifSFB@referee
 
\else
 
\fi
\if@twocolumn
  \def\@normalsize{\@setsize\normalsize{11pt}\ixpt\@ixpt
   \abovedisplayskip 6pt plus 2pt minus 2pt
   \belowdisplayskip \abovedisplayskip
   \abovedisplayshortskip 6pt plus 2pt
   \belowdisplayshortskip \abovedisplayshortskip
   \let\@listi\@listI}
 \else
 \ifSFB@referee
  \def\@normalsize{\@setsize\normalsize{14pt}\xiipt\@xiipt
   \abovedisplayskip 4pt plus 1pt minus 1pt
   \belowdisplayskip \abovedisplayskip
   \abovedisplayshortskip 4pt plus 1pt
   \belowdisplayshortskip \abovedisplayshortskip
   \let\@listi\@listI}
 \else
  \def\@normalsize{\@setsize\normalsize{12pt}\ixpt\@ixpt
   \abovedisplayskip 4pt plus 1pt minus 1pt
   \belowdisplayskip \abovedisplayskip
   \abovedisplayshortskip 4pt plus 1pt
   \belowdisplayshortskip \abovedisplayshortskip
   \let\@listi\@listI}
 \fi
\fi
\def\small{\@setsize\small{10pt}\viiipt\@viiipt
 \abovedisplayskip 4pt plus 1pt minus 1pt
 \belowdisplayskip \abovedisplayskip
 \abovedisplayshortskip 4pt plus 1pt
 \belowdisplayshortskip \abovedisplayshortskip
 \def\@listi{\leftmargin\leftmargini
  \topsep 2pt plus 1pt minus 1pt
  \parsep \z@
  \itemsep 2pt}}
\def\footnotesize{\@setsize\footnotesize{10pt}\viiipt\@viiipt
 \abovedisplayskip 4pt plus 1pt minus 1pt
 \belowdisplayskip \abovedisplayskip
 \abovedisplayshortskip 4pt plus 1pt
 \belowdisplayshortskip \abovedisplayshortskip
 \def\@listi{\leftmargin\leftmargini
  \topsep 2pt plus 1pt minus 1pt
  \parsep \z@
  \itemsep 2pt}}
\def\scriptsize{\@setsize\scriptsize{8pt}\viipt\@viipt}
\def\tiny{\@setsize\tiny{6pt}\vpt\@vpt}
\if@twocolumn
  \def\large{\@setsize\large{11pt}\xpt\@xpt}
 \else
  \def\large{\@setsize\large{12pt}\xpt\@xpt}
 \fi
\def\Large{\@setsize\Large{14pt}\xiipt\@xiipt}
\def\LARGE{\@setsize\LARGE{17pt}\xivpt\@xivpt}
\def\huge{\@setsize\huge{20pt}\xviipt\@xviipt}
\def\Huge{\@setsize\huge{25pt}\xxpt\@xxpt}
\normalsize

\if@twocolumn
\else
 \ifSFB@referee
  \oddsidemargin \z@  \evensidemargin \z@
 \else
 \fi
\fi
 \topmargin \z@
\newdimen\SFB@measure
\textwidth \SFB@measure
\ifSFB@landscape
 \textwidth \textheight
 \textheight \SFB@measure
\fi
\ifSFB@referee
\fi
\skip\footins 19.5pt plus 12pt minus 1pt

\parskip \z@ plus .1pt
\@beginparpenalty -\@lowpenalty
\@endparpenalty -\@lowpenalty
\@itempenalty -\@lowpenalty
\newcounter{part}
\newcounter {section}
\newcounter {subsection}[section]
\newcounter {subsubsection}[subsection]
\newcounter {paragraph}[subsubsection]
\newcounter {subparagraph}[paragraph]
\def\thepart          {\arabic{part}}
\def\thesection       {\arabic{section}}

\def\part{\par \addvspace{4ex}\@afterindentfalse
 \secdef\@part\@spart}
\def\@part[#1]#2{\ifnum \c@secnumdepth >\m@ne
  \refstepcounter{part}
  \addcontentsline{toc}{part}{Part \thepart: #1}
 \else \addcontentsline{toc}{part}{#1}
 \fi
 {\parindent 0pt \raggedright
  \ifnum \c@secnumdepth >\m@ne
   \large\rm PART
   \ifcase\thepart \or ONE \or TWO \or THREE \or FOUR \or FIVE
    \or SIX \or SEVEN \or EIGHT \or NINE \or TEN \else \fi
   \par \nobreak
  \fi
  \LARGE \rm #2 \markboth{}{}\par }
 \nobreak \vskip 3ex \@afterheading}
\def\@spart#1{{\parindent 0pt \raggedright
  \LARGE \rm #1\par}
 \nobreak \vskip 3ex \@afterheading}

\def\section{\@startsection {section}{1}{\z@}
 {-24pt plus -12pt minus -1pt}
 {6pt}
 {\SFB@hangraggedright\normalsize\bf}}
\def\subsection{\@startsection{subsection}{2}{\z@}
 {-18pt plus -9pt minus -1pt}
 {6pt}
 {\SFB@hangraggedright\large\bf}}
\def\subsubsection{\@startsection{subsubsection}{3}{\z@}
 {-18pt plus -9pt minus -1pt}
 {6pt}
 {\SFB@hangraggedright\normalsize\it}}
\def\paragraph{\@startsection{paragraph}{4}{\z@}
 {12pt plus 2.25pt minus 1pt}{-0.5em}{\normalsize\bf}}
\def\subparagraph{\@startsection{subparagraph}{5}{\parindent}
 {12pt plus 2.25pt minus 1pt}{-0.5em}{\normalsize\it}}
\def\SFB@hangraggedright{\rightskip\@flushglue \let\\=\newline}
\def\@sect#1#2#3#4#5#6[#7]#8{%
 \ifnum #2>\c@secnumdepth
  \def\@svsec{}%
 \else
  \refstepcounter{#1}
  \ifnum #2=\@ne
   \ifSFB@appendix \edef\@svsec{}%
             \else \edef\@svsec{\csname the#1\endcsname\hskip 1em}%
   \fi
  \else \edef\@svsec{\csname the#1\endcsname\hskip 1em}%
  \fi
 \fi
 \@tempskipa #5\relax
 \ifdim \@tempskipa>\z@
  \begingroup #6\relax
   \ifnum #2=\@ne
    \ifSFB@appendix
     \@hangfrom{\hskip #3\relax\@svsec}{\interlinepenalty \@M
      APPENDIX \csname the#1\endcsname:\hskip 0.5em\uppercase{#8}\par}%
    \else
     \@hangfrom{\hskip #3\relax\@svsec}{\interlinepenalty \@M
      \uppercase{#8}\par}%
    \fi
   \else
    \@hangfrom{\hskip #3\relax\@svsec}{\interlinepenalty \@M #8\par}%
   \fi
  \endgroup
  \csname #1mark\endcsname{#7}%
  \addcontentsline{toc}{#1}{\ifnum #2>\c@secnumdepth \else
   \protect\numberline{\csname the#1\endcsname}\fi #7}%
 \else
  \def\@svsechd{#6\hskip #3\@svsec \ifnum #2=\@ne\uppercase{#8}\else #8\fi
  \csname #1mark\endcsname{#7}
  \addcontentsline{toc}{#1}{\ifnum #2>\c@secnumdepth \else
   \protect\numberline{\csname the#1\endcsname}\fi#7}}%
 \fi
 \@xsect{#5}}

\newif\ifSFB@appendix
\def\appendix{\par
 \SFB@appendixtrue
 \setcounter{section}{0}
 \def\thesection{A\arabic{section}}
 \setcounter{equation}{0}
 \def\theequation{A\arabic{equation}}
 \setcounter{figure}{0}
 \def\thefigure{A\@arabic\c@figure}
 \setcounter{table}{0}
 \def\thetable{A\@arabic\c@table}
}

\newskip\@indentskip
\newskip\smallindent
\newskip\@footindent
\newskip\@leftskip
\@footindent=\smallindent
\@leftskip=\z@

\leftmargini   \@indentskip
\leftmargin\leftmargini
\labelwidth\leftmargini\advance\labelwidth-\labelsep
\def\makeRRlabel#1{\hss\llap{#1}}
\def\@listI{\leftmargin\leftmargini
 \parsep \z@
 \topsep 6pt plus 1pt minus 1pt
 \itemsep \z@ plus .1pt
}
\let\@listi\@listI
\@listi
\def\@listii{\leftmargin\leftmarginii
 \labelwidth\leftmarginii\advance\labelwidth-\labelsep
 \topsep 6pt plus 1pt minus 1pt
 \parsep \z@
 \itemsep \z@ plus .1pt
}
\def\@listiii{\leftmargin\leftmarginiii
 \labelwidth\leftmarginiii\advance\labelwidth-\labelsep
 \topsep 6pt plus 1pt minus 1pt
 \parsep \z@
 \partopsep \z@
 \itemsep \topsep
}
\def\@listiv{\leftmargin\leftmarginiv
 \labelwidth\leftmarginiv\advance\labelwidth-\labelsep
}
\def\@listv{\leftmargin\leftmarginv
 \labelwidth\leftmarginv\advance\labelwidth-\labelsep
}
\def\@listvi{\leftmargin\leftmarginvi
 \labelwidth\leftmarginvi\advance\labelwidth-\labelsep
}
\def\itemize{\ifnum \@itemdepth >3 \@toodeep
  \else \advance\@itemdepth \@ne
   \edef\@itemitem{labelitem\romannumeral\the\@itemdepth}%
   \list{\csname\@itemitem\endcsname}%
    {\let\makelabel\makeRRlabel}%
  \fi}

\def\enumerate{\ifnum \@enumdepth >3 \@toodeep \else
  \advance\@enumdepth \@ne
  \edef\@enumctr{enum\romannumeral\the\@enumdepth}%
 \fi
 \@ifnextchar [{\@enumeratetwo}{\@enumerateone}%
}
\def\@enumeratetwo[#1]{%
 \list{\csname label\@enumctr\endcsname}%
  {\settowidth\labelwidth{[#1]}
   \leftmargin\labelwidth \advance\leftmargin\labelsep
   \usecounter{\@enumctr}
   \let\makelabel\makeRRlabel}
}
\def\@enumerateone{%
 \list{\csname label\@enumctr\endcsname}%
  {\usecounter{\@enumctr}
   \let\makelabel\makeRRlabel}}

\def\theenumi{(\roman{enumi})}

\def\theenumii{(\alph{enumii})}
\def\p@enumii{\theenumi}

\def\theenumiii{(\arabic{enumiii})}
\def\p@enumiii{\theenumi(\theenumii)}

\def\p@enumiv{\p@enumiii\theenumiii}
\def\description{\list{}{\labelwidth\z@ \itemindent-\leftmargin
  \leftmargin 1em
  \itemindent-1em
}}

\def\verse{\let\\=\@centercr
 \list{}{\itemsep\z@
  \itemindent -\@indentskip
  \listparindent \itemindent
  \rightmargin\leftmargin
  \advance\leftmargin \@indentskip}\item[]}

\def\quotation{\list{}{\listparindent \smallindent
  \leftmargin\z@\rightmargin\leftmargin
  \parsep 0pt plus 1pt}\item[]\small}

\def\quote{\list{}{\leftmargin\z@\rightmargin\leftmargin}\item[]\small}

\def\@begintheorem#1#2{\rm \trivlist \item[\hskip \labelsep{\bf #1\ #2.}]}
\def\@opargbegintheorem#1#2#3{\rm \trivlist
  \item[\hskip \labelsep{\bf #1\ #2.\ (#3)}]}
\def\@endtheorem{\endtrivlist}
\@namedef{proof*}{\rm \trivlist \item[\hskip \labelsep{\it Proof.}]}
\@namedef{endproof*}{\endtrivlist}

\def\titlepage{\@restonecolfalse\if@twocolumn\@restonecoltrue\onecolumn
  \else \newpage \fi \thispagestyle{empty}\c@page\z@}
\def\endtitlepage{\if@restonecol\twocolumn \else \newpage \fi}

\def\tabular{\def\@halignto{}
 \def\hline{\noalign{\ifnum0=`}\fi
  \vskip 3pt
  \hrule \@height \arrayrulewidth
  \vskip 3pt
  \futurelet \@tempa\@xhline}
 \def\fullhline{\noalign{\ifnum0=`}\fi
  \vskip 3pt
  \hrule \@height \arrayrulewidth
  \vskip 3pt
  \futurelet \@tempa\@xhline}
 \def\@xhline{\ifx\@tempa\hline
   \vskip -6pt
   \vskip \doublerulesep
  \fi
  \ifnum0=`{\fi}}
  \def\@arrayrule{\@addtopreamble{\hskip -.5\arrayrulewidth
                                  \hskip .5\arrayrulewidth}}
\@tabular
}
\tabbingsep \labelsep

\skip\@mpfootins = \skip\footins

\fboxrule = \arrayrulewidth

\def\maketitle{\par
 \begingroup
  \if@twocolumn
   \twocolumn[\vspace*{17pt}\@maketitle]
  \else
   \newpage
   \global\@topnum\z@
   \@maketitle
  \fi
  \thispagestyle{titlepage}
 \endgroup
 \let\maketitle\relax
 \let\@maketitle\relax
 \gdef\@author{}
 \gdef\@title{}
 \let\thanks\relax
}
\def\and{\end{author@tabular}\vskip 6pt\par
 \begin{author@tabular}[t]{@{}l@{}}}
\def\@maketitle{\newpage
 \vspace*{7pt}
 {\raggedright \sloppy
  {\huge \bf \@title \par}
  \vskip 23pt
  {\LARGE
   \begin{author@tabular}[t]{@{}l@{}}\@author
   \end{author@tabular}\par}
  \vskip 22pt
 }
 \par\noindent
 {\small \@date \par}
 \vskip 22pt
}
\def\abstract{\if@twocolumn
  \start@SFBbox\@abstract
 \else
  \@abstract
 \fi}
\def\endabstract{\if@twocolumn
   \endlist\finish@SFBbox
 \else
  \endlist
 \fi}
\def\@abstract{\list{}{\leftmargin 10.5pc\rightmargin\z@
  \parsep 0pt plus 1pt}\item[]\normalsize{\bf ABSTRACT}\\\large} 
\newif\ifSFB@keywords
\def\keywords{\if@twocolumn
  \start@SFBbox\@keywords
 \else
  \@keywords
 \fi
}
\def\@keywords{\list{}{\leftmargin 10.5pc\rightmargin\z@
  \parsep 0pt plus 1pt}\item[]\large{\bf Key words: }}
\def\endkeywords{\if@twocolumn
  \endlist\addvspace{37pt}\finish@SFBbox
 \else
  \endlist
 \fi
 \@thanks
 \gdef\@thanks{}
 \SFB@keywordstrue
}
\def\nokeywords{\ifSFB@keywords\else
 \if@twocolumn \start@SFBbox\addvspace{37pt}\finish@SFBbox \fi
 \@thanks
 \gdef\@thanks{}\fi
}

\def\author@tabular{\def\@halignto{}\@authortable}
\let\endauthor@tabular=\endtabular
\def\author@tabcrone{{\ifnum0=`}\fi\@xtabularcr[-7pt]\small\it
 \let\\=\author@tabcrtwo\ignorespaces}
\def\author@tabcrtwo{{\ifnum0=`}\fi\@xtabularcr[-7pt]\small\it
 \let\\=\author@tabcrtwo\ignorespaces}
\def\@authortable{\leavevmode \hbox \bgroup $\let\@acol\@tabacol
 \let\@classz\@tabclassz \let\@classiv\@tabclassiv
 \let\\=\author@tabcrone \ignorespaces \@tabarray}

\def\start@SFBbox{\@next\@currbox\@freelist{}{}%
 \global\setbox\@currbox
 \vbox\bgroup
  \hsize \textwidth
  \@parboxrestore
}
\def\finish@SFBbox{\par\vskip -\dbltextfloatsep
  \egroup
  \global\count\@currbox\tw@
  \global\@dbltopnum\@ne
  \global\@dbltoproom\maxdimen\@addtodblcol
  \global\vsize\@colht
  \global\@colroom\@colht
}

\mark{{}{}}
\gdef\@author{\mbox{}}
\def\author{\@ifnextchar [{\@authortwo}{\@authorone}}
\def\@authortwo[#1]#2{\gdef\@author{#2}\gdef\@shortauthor{#1}}
\def\@authorone#1{\gdef\@author{#1}\gdef\@shortauthor{#1}}
\gdef\@shortauthor{}
\gdef\@title{\mbox{}}
\def\title{\@ifnextchar [{\@titletwo}{\@titleone}}
\def\@titletwo[#1]#2{\gdef\@title{#2}\gdef\@shorttitle{#1}}
\def\@titleone#1{\gdef\@title{#1}\gdef\@shorttitle{#1}}
\gdef\@shorttitle{}
\def\volume#1{\gdef\@volume{#1}}
\gdef\@volume{000}
\def\microfiche#1{\gdef\@microfiche{#1}}
\gdef\@microfiche{}
\def\pagerange#1{\gdef\@pagerange{#1}}
\gdef\@pagerange{000--000}
\def\journal#1{\gdef\@journal{#1}}
\gdef\@journal{{Mon.\ Not.\ R.\ Astron.\ Soc.} {\bf \@volume}, \@pagerange\
  (\number\year) \@microfiche}
\def\ps@headings{\let\@mkboth\markboth
 \def\@oddhead{\Large \hfill \it \@shorttitle \hspace{1.5em}\rm \thepage}
 \def\@oddfoot{}
 \def\@evenhead{\Large \thepage \hspace{1.5em}\it \@shortauthor \hfill}
 \def\@evenfoot{}
 \def\sectionmark##1{\markboth{##1}{}}
 \def\subsectionmark##1{\markright{##1}}}
\def\ps@myheadings{\let\@mkboth\@gobbletwo
 \def\@oddhead{\Large \it \rightmark \hfill \rm \thepage}
 \def\@oddfoot{}
 \def\@evenhead{\Large \it \leftmark \hfill \rm \thepage}
 \def\@evenfoot{}
 \def\sectionmark##1{}
 \def\subsectionmark##1{}}
\def\ps@titlepage{\let\@mkboth\@gobbletwo
 \def\@oddhead{\footnotesize\@journal\hfill}
 \def\@oddfoot{}
 \def\@evenhead{\footnotesize\@journal\hfill}
 \def\@evenfoot{}
 \def\sectionmark##1{}
 \def\subsectionmark##1{}}

\def\@pnumwidth{1.55em}
\def\@tocrmarg {2.55em}
\def\@dotsep{4.5}
\def\@undottedtocline#1#2#3#4#5{\ifnum #1>\c@tocdepth
 \else
  \vskip \z@ plus .2pt
  {\hangindent #2\relax
   \rightskip \@tocrmarg \parfillskip -\rightskip
   \parindent #2\relax \@afterindenttrue
   \interlinepenalty\@M \leavevmode
   \@tempdima #3\relax #4\nobreak \hfill \nobreak
   \hbox to\@pnumwidth{\hfil\rm \ }\par}\fi}
\def\tableofcontents{\@restonecolfalse
 \if@twocolumn\@restonecoltrue\onecolumn\fi
 \section*{CONTENTS} \@starttoc{toc}
 \if@restonecol\twocolumn\fi \par\vspace{12pt}}
\def\l@part#1#2{\addpenalty{-\@highpenalty}
 \addvspace{2.25em plus 1pt}
 \begingroup
  \parindent \z@ \rightskip \@pnumwidth
  \parfillskip -\@pnumwidth
  {\normalsize\rm
   \leavevmode \hspace*{3pc}
   #1\hfil \hbox to\@pnumwidth{\hss \ }}\par
   \nobreak \global\@nobreaktrue
   \everypar{\global\@nobreakfalse\everypar{}}\endgroup}
\def\l@section#1#2{\addpenalty{\@secpenalty}
 \@tempdima 1.5em
 \begingroup
  \parindent \z@ \rightskip \@pnumwidth
  \parfillskip -\@pnumwidth \rm \leavevmode
  \advance\leftskip\@tempdima \hskip -\leftskip
  #1\nobreak\hfil \nobreak\hbox to\@pnumwidth{\hss \ }\par
 \endgroup}
\def\l@subsection{\@undottedtocline{2}{1.5em}{2.3em}}
\def\l@subsubsection{\@undottedtocline{3}{3.8em}{3.2em}}
\def\l@paragraph{\@undottedtocline{4}{7.0em}{4.1em}}
\def\l@subparagraph{\@undottedtocline{5}{10em}{5em}}
\def\listoffigures{\@restonecolfalse
 \if@twocolumn\@restonecoltrue\onecolumn\fi
 \section*{LIST OF FIGURES\@mkboth{LIST OF FIGURES}{LIST OF FIGURES}}
 \@starttoc{lof} \if@restonecol\twocolumn\fi}
\def\l@figure{\@undottedtocline{1}{1.5em}{2.3em}}
\def\listoftables{\@restonecolfalse
 \if@twocolumn\@restonecoltrue\onecolumn\fi
 \section*{LIST OF TABLES\@mkboth{LIST OF TABLES}{LIST OF TABLES}}
 \@starttoc{lot} \if@restonecol\twocolumn\fi}
\let\l@table\l@figure

\def\thebibliography#1{\section*{REFERENCES}
 \addcontentsline{toc}{section}{REFERENCES}
 \list{}{\labelwidth\z@
         \leftmargin 1.5em
	 \itemsep \z@
	 \itemindent-\leftmargin}
 \small\raggedright
 \parindent\z@
 \parskip\z@ plus .1pt\relax
 \def\newblock{\hskip .11em plus .33em minus .07em}
 \sloppy\clubpenalty4000\widowpenalty4000
 \sfcode`\.=1000\relax
}

\def\@biblabel#1{\hspace*{\labelsep}[#1]}

\newif\if@restonecol
\def\theindex{\section*{INDEX}
 \addcontentsline{toc}{section}{INDEX}
 \footnotesize \parindent\z@ \parskip\z@ plus .1pt\relax
 \let\item\@idxitem}
\def\@idxitem{\par\hangindent 1em}

\def\endtheindex{\if@restonecol\onecolumn\else\clearpage\fi}

\def\footnoterule{\kern-3\p@ \hrule width 12pc height \z@ \kern 3\p@}

\def\@fnsymbol#1{\ifcase#1\or \mbox{$\star$}\or \dagger\or \ddagger\or
   \S \or \P \or \|\or **\or \dagger\dagger
   \or \ddagger\ddagger\or \S\S\or \P\P\or \|\|\else ***
   \fi\relax}

\long\def\@makefntext#1{\parindent 1em\noindent
  $^{\@thefnmark}$\hspace{4pt}#1}

\newcounter{table}
\def\thetable{\@arabic\c@table}
\def\fps@table{tbp}
\def\ftype@table{1}
\def\ext@table{lot}
\def\fnum@table{Table \thetable}
\def\table{\let\@makecaption=\SFB@maketablecaption\@float{table}}
\let\endtable\end@float
\@namedef{table*}{\let\@makecaption=\SFB@maketablecaption\@dblfloat{table}}
\@namedef{endtable*}{\end@dblfloat}

\newcounter{figure}
\def\thefigure{\@arabic\c@figure}
\def\fps@figure{tbp}
\def\ftype@figure{2}
\def\ext@figure{lof}
\def\fnum@figure{Figure \thefigure}
\def\figure{\let\@makecaption=\SFB@makefigurecaption\@float{figure}}
\let\endfigure\end@float
\@namedef{figure*}{\let\@makecaption=\SFB@makefigurecaption\@dblfloat{figure}}
\@namedef{endfigure*}{\end@dblfloat}

\long\def\SFB@makefigurecaption#1#2{\vskip 6pt
 \setbox\@tempboxa\hbox{\small{\bf #1.} #2}
 \ifdim \wd\@tempboxa >\hsize
  \small{\bf #1.} #2\par
 \else
  \hbox to\hsize{\hfil\box\@tempboxa\hfil}
 \fi
 \vskip 6pt
}
\long\def\SFB@maketablecaption#1#2{\vskip 6pt
 \setbox\@tempboxa\hbox{\small{\bf #1.} #2}
 \ifdim \wd\@tempboxa >\hsize
  \small{\bf #1.} #2\par
 \else
  \hbox to\hsize{\box\@tempboxa\hfill}
 \fi
 \vskip 6pt
}

\def\caption{\@ifstar{\SFB@caption\@captype}%
 {\refstepcounter\@captype \@dblarg{\@caption\@captype}}%
}
\long\def\SFB@caption#1#2{
 \begingroup
  \@parboxrestore
  \normalsize
  \@makecaption{\csname fnum@#1\endcsname}{\ignorespaces #2}\par
 \endgroup}

\def\@cite#1#2{(#1\if@tempswa , #2\fi)}
\def\@biblabel#1{}

\newlength{\bibhang}
\def\@citex[#1]#2{\if@filesw\immediate\write\@auxout{\string\citation{#2}}\fi
  \def\@citea{}\@cite{\@for\@citeb:=#2\do
    {\@citea\def\@citea{; }\@ifundefined
       {b@\@citeb}{{\bf ?}\@warning
       {Citation `\@citeb' on page \thepage \space undefined}}%
{\csname b@\@citeb\endcsname}}}{#1}}
\let\@internalcite\cite
\def\cite{\def\citename##1{##1}\@internalcite}
\def\shortcite{\def\citename##1{}\@internalcite}

\def\[{\relax\ifmmode\@badmath\else\begin{trivlist}\item[]\leavevmode
  \hbox to\linewidth\bgroup$
  \displaystyle
  \hskip\mathindent\bgroup\fi}

\def\]{\relax\ifmmode \egroup $\hfil
       \egroup \end{trivlist}\else \@badmath \fi}

\def\equation{\refstepcounter{equation}\trivlist \item[]\leavevmode
  \hbox to\linewidth\bgroup $
  \displaystyle
\hskip\mathindent}

\def\endequation{$\hfil
           \displaywidth\linewidth\@eqnnum\egroup \endtrivlist}

\def\eqnarray{\stepcounter{equation}\let\@currentlabel=\theequation
\global\@eqnswtrue
\global\@eqcnt\z@\tabskip\mathindent\let\\=\@eqncr
\abovedisplayskip\topsep\ifvmode\advance\abovedisplayskip\partopsep\fi
\belowdisplayskip\abovedisplayskip
\belowdisplayshortskip\abovedisplayskip
\abovedisplayshortskip\abovedisplayskip
$$\halign
to \linewidth\bgroup\@eqnsel\hskip\@centering$\displaystyle\tabskip\z@
  {##}$&\global\@eqcnt\@ne \hskip 2\arraycolsep \hfil${##}$\hfil
  &\global\@eqcnt\tw@ \hskip 2\arraycolsep $\displaystyle{##}$\hfil
   \tabskip\@centering&\llap{##}\tabskip\z@\cr}

\def\endeqnarray{\@@eqncr\egroup
 \global\advance\c@equation\m@ne$$\global\@ignoretrue}

\newdimen\mathindent
\mathindent = \z@

\def\today{\number\day\ \ifcase\month\or
  January\or February\or March\or April\or May\or June\or
  July\or August\or September\or October\or November\or December
 \fi \ \number\year}

\ps@headings
\ifSFB@galley
 \ps@empty
\fi
\ifSFB@referee
\fi
\if@twocolumn
 \twocolumn
\else
 \onecolumn
\fi
\documentstyle
{mn}

\def\etal{et~al.\/}
\def\g{$\gamma$}

\newbox\grsign \setbox\grsign=\hbox{$>$}
\newdimen\grdimen \grdimen=\ht\grsign
\newbox\laxbox \newbox\gaxbox
\setbox\gaxbox=\hbox{\raise.5ex\hbox{$>$}\llap
     {\lower.5ex\hbox{$\sim$}}}\ht1=\grdimen\dp1=0pt
\setbox\laxbox=\hbox{\raise.5ex\hbox{$<$}\llap
     {\lower.5ex\hbox{$\sim$}}}\ht2=\grdimen\dp2=0pt
\def\gax{\mathrel{\copy\gaxbox}}
\def\lax{\mathrel{\copy\laxbox}}
\def\simless{\lax}
\def\simgreat{\gax}

\def\Cbarsix{(\bar{C}^{64})_{\rm max}}

\def\Cbarten{(\bar{C}^{1024})_{\rm max}}

\begin{document}

\title[Evidence that $\gamma$-ray burst sources repeat]
      {Evidence that $\gamma$-ray burst sources repeat}

\author
[J. M. Quashnock and D. Q. Lamb]
	{J. M. Quashnock and D. Q. Lamb \\
	Department of Astronomy and Astrophysics, University of Chicago,
	Chicago, IL 60637}

\date{Accepted 1993 October 15.  Received 1993 September 20; in
original form 1993 July 6.}

\maketitle

\begin{abstract}

We investigate clustering in the angular distribution of the 260
$\gamma$-ray bursts in the publicly available BATSE catalogue, using a
nearest neighbour analysis and the measures of burst brightness $B$ and
short time scale variability $V$ which we introduced earlier.  We find
that while all 260 bursts are only modestly clustered (Q-value = $1.8
\times 10^{-2}$), the 202 bursts in this sample for which the
statistical error in their locations is $<9^\circ$ are significantly
clustered on an angular scale $\approx 5^\circ$ (Q-value = $2.5 \times
10^{-4}$, taking into acoount having chosen the cutoff in the
statistical error).  We also find a significant correlation between
bright type I bursts and faint type I and type II bursts on an angular
scale $\approx 5^\circ$ (Q-value = $4.0 \times 10^{-3}$).  This angular
scale is smaller than the typical (statistical plus systematic) error
in burst locations of $6.8^\circ$, suggesting multiple recurrences from
individual sources.  We conclude that ``classical'' $\gamma$-ray burst
sources repeat on a time scale of months, and that many faint type I
and II bursts come from the sources of bright type I bursts.

\end{abstract}


\section{Introduction}

The ``soft $\gamma$-ray repeaters'' SGR 0526-66, SGR 1806-20, and SGR
1900+14 have been observed to produce more than 100 bursts.  The bursts
typically have short durations ($t_{\rm dur} \approx 250$ ms) and soft
spectra (characteristic energies $E \approx 30$ keV), and are
sufficiently different from ``classical'' $\gamma$-ray bursts that
these three sources are thought to constitute a separate class (see,
e.g., Higdon and Lingenfelter 1990, Harding 1992, Hurley 1992).  In
contrast, no source of a ``classical'' $\gamma$-ray burst has been
found to repeat during nearly twenty five years of observations.
Assuming that $\gamma$-ray bursts are standard candles, Schaefer and
Cline (1985) estimate a recurrence time scale $t_{\rm recur} \simgreat
10$ yrs from analysis of the locations of 89 bright bursts with
relatively small error boxes, and Atteia et al. (1987) derive a
$3\sigma$ lower limit of $t_{\rm recur} > 8$ yrs from analysis of the
locations of 84 bright bursts with error boxes determined by the
Interplanetary Network.  However, both limits become $t_{\rm recur}
\simgreat$ a few months if the burst luminosity function is broad.

The BATSE catalogue (Fishman \etal\ 1993) is the largest homogeneous
sample of \g-ray bursts ever assembled and, as such, constitutes a
unique resource for studying the angular distribution of bursts.
Analysis of clustering in the angular distribution of the bursts can
indicate whether or not individual burst sources repeat.  Here we
investigate such clustering using a nearest neighbour analysis and the
measures of burst brightness $B = \Cbarten$ and short time scale
($\simless$ 0.3 s) variability $V = \Cbarsix / \Cbarten$, which we
introduced earlier (Lamb, Graziani, and Smith 1993).  The quantities
$\Cbarsix$ and $\Cbarten$ are the expected maximum number of counts in
64 ms and in 1024 ms, respectively.

\section{Analysis}

In earlier studies (Lamb, Graziani, and Smith 1993; Lamb and Graziani
1993a,b), we presented evidence for two distinct morphological classes
of $\gamma$-ray bursts (see also Kouveliotou \etal\ 1993).  Type I
bursts (comprising $\approx$ 80\% of the bursts) are smoother ($\log V
\le -0.8$) on short time scales ($\simless 0.3$ s), both faint and
bright, longer, and have softer spectra.  Type II bursts (comprising
$\approx$ 20\% of the bursts) are more variable ($\log V > -0.8$) on
short time scales ($\simless 0.3$ s), faint, shorter, and have harder
spectra.

Fig.~1 shows the locations on the sky of the 260 bursts in the publicly
available BATSE catalogue (Fishman \etal\ 1993).  Because of the
relatively small number of bursts, weak clustering is detectable only
on angular scales $\simgreat 10^\circ$ (see, e.g., Lamb and Quashnock
1993).  Yet the eye is drawn to a number of places on the sky where two
or more bursts lie very close together, suggesting strong (nonlinear)
clustering and hence multiple recurrences from individual sources.

Standard analysis techniques for investigating the clustering of
objects on the sky include decomposition of the distribution of objects
in terms of spherical harmonics $Y^l_m$ (Hartmann \& Epstein 1989) and
calculation of the two-point angular correlation function $w(\theta)$
(Hartmann \& Blumenthal 1989); in fact, $w(\theta)$ can be written as a
sum of Legendre polynomials times the coefficients of the spherical
harmonic decomposition (Peebles 1980).

However, decomposition into spherical harmonics is relatively poor at
detecting sharp features (such as nonlinear clustering) because the
power is spread over many high harmonics.  Calculation of the two-point
angular correlation function is also relatively poor at detecting
nonlinear clustering because it does not include higher order
correlations; in addition, it spreads the clustering over a larger
angular scale because it includes {\it all} the angular separations
between bursts in the cluster, averaged over all pairs of bursts.

Here we consider a related statistic, the nearest neighbour separation,
which is particularly sensitive to nonlinear clustering and thus to
multiple recurrences from individual sources.  Every burst has a
nearest neighbour; indeed, for a random distribution, the probability
that a nearest neighbour is found within an angle $\theta$ from a
randomly chosen object is (Scott and Tout 1989)
\begin{equation}
P(\theta) = 1 - [(1+\cos \theta)/2]^N \; ,
\end{equation}
where $N$ is the total number of {\it neighbouring} bursts on the sky.
Monte Carlo calculations show that for $N \approx 200$ the effect of
the BATSE sky exposure map is small [${\cal O}(10^{-3})$] on the
angular scales ($\theta \simless 10^\circ$) of interest in this work,
and we therefore do not include it in $P(\theta)$.  We note that for a
random distribution uncertainties in the burst locations have no effect
on $P(\theta)$.

We investigate the clustering of $\gamma$-ray bursts as a function of
burst brightness $B$ by calculating the cumulative distribution of
nearest neighbour separations {\it within} the sample of all 260
bursts, all 201 bursts for which $B$ and $V$ exist, all type I bursts,
and faint ($B \le 1900$ counts) type I bursts, as well as {\it between}
various brightness sub-samples of bursts.  A condition for the validity
of the Kolmogorov-Smirnov test is that all the measurements be
independent.  This condition is not satisfied for nearest neighbour
separations {\it within} a sample, since the nearest neighbour of a
given burst may often have the given burst as its own nearest
neighbour.  In this case, we evaluate the significance of the largest
deviation $D$ of the cumulative distribution of nearest neighbour
separations from that expected for a random distribution with the same
number of bursts using Monte Carlo simulations.  We determine the
fraction of simulations which exhibit positive {\it or negative} values
of $D$ that equal or exceed the magnitude of the observed $D$.  We
evaluate the significance of the largest deviation $D$ of the
cumulative distribution of nearest neighbour separations {\it between}
two samples from that expected for a random distribution with the same
number of bursts using the Kolmogorov-Smirnov test (Press et al. 1986),
since in this case the nearest neighbour separations are independent.

\section{Results}

Fig.~ 2 (top panel) shows the cumulative nearest neighbour
distributions for all bursts in the publicly available BATSE catalogue
(heavy histogram).  Fig.~2 (bottom panel) shows the cumulative nearest
neighbour distributions for various sub-samples.  Also shown are the
expected distributions, given by equation (1), for random distributions
with the same total number of bursts.  Table 1 gives the significance
of the maximum deviations $D$ of each from that expected for a random
distribution with the same number of bursts.

Fig.~2 (top panel) shows that the sample of all 260 bursts has more
nearest neighbours at small separations than expected for a random
distribution (Q-value = $1.8 \times 10^{-2}$).  If this excess is due
to repeating, one expects its significance to increase as bursts with
large statistical errors in their locations are removed from the
sample, and then to decrease as bursts with smaller statistical errors
are removed.  This is the case.  The maximum in the significance is
$1.1 \times 10^{-4}$ and occurs at a cutoff of $\approx 9^\circ$ in the
statistical error.  Fig.~2 (top panel) shows the resulting cumulative
nearest neighbour distribution (light histogram).  The corresponding
total (statistical plus systematic) error of $\approx 10^\circ$ is
larger than the angular scale $\approx 5^\circ$ of the excess of
nearest neighbour separations.  This suggests multiple recurrences of
individual sources, since bursts with total errors larger than the
angular scale of the excess then contribute importantly to the signal
(see below).

Using Monte Carlo simulations, we evaluate the amount by which the
latter significance should be reduced in order to take into account
having chosen a cutoff in the statistical error.  In order to find the
maximum significance, we considered cutoffs of $8^\circ$, $9^\circ$,
and $10^\circ$.  We therefore calculate the fraction of simulations of
211 bursts (corresponding to the sample size for a cutoff of $10^
\circ$) which exhibit positive {\it or negative} values of $D$ that
equal or exceed the observed significance of $1.1 \times 10^{-4}$.
Next we remove nine bursts at random from each of the simulations of
211 bursts which did {\it not} meet the above criterion (yielding
simulations of 202 bursts, which corresponds to the sample size for a
cutoff of $9^\circ$), and repeat the calculation.  Finally we remove
fifteen bursts at random from each of the simulations of 202 bursts
which did {\it not} meet the above criterion (yielding simiulations of
187 bursts, which corresponds to the sample size for a cutoff of
$8^\circ$), and repeat the calculation again.  The sum of these
fractions gives the probability by chance of a random distribution of
260 bursts exhibiting positive {\it or negative} values of $D$ that
equal or exceed a significance of $1.1 \times 10^{-4}$, having
considered three cutoffs.  We obtain a significance of $2.5 \times
10^{-4}$ in this way.

Fig.~2 (bottom panel) shows that faint ($B \le 1900$ counts) type I
bursts have more nearest neighbours at small separations than expected
for a random distribution (Q-value = $1.3 \times 10^{-2}$), while all
type I bursts and all (type I and II) bursts for which $B$ and $V$
exist have significantly more than expected (Q-values = $4.8 \times
10^{-3}$ and $2.1 \times 10^{-3}$, respectively).  Retaining only the
159 bursts in the last sample for which the statistical error in their
burst locations is $<9^\circ$ again increases the significance of the
clustering (Q-value = $7.9 \times 10^{-4}$).  We do not evaluate the
amount by which the last significance should be reduced in order to
take into account having chosen a cutoff, because it differs only
modestly from the significance for the full sample.

The increase in significance going from the sample of faint type I
bursts to all type I bursts implies that faint type I bursts are
correlated with bright type I bursts.  Similarly, the increase in
significance going from the sample of all type I bursts to all bursts
implies that type II bursts are correlated with type I bursts.  The
latter implication is strengthened by inspection of the samples that
contain type II bursts; this reveals that many type II bursts
contribute to the excess in the nearest neighbour distribution at small
angular separations in these samples.

The typical systematic error $\theta_{\rm sys}$ in the locations of the
bursts in the publicly available BATSE catalogue is $4^\circ$; the
median statistical errors $\theta_{\rm stat}$ for faint type I bursts,
all type I bursts, and all bursts are 5.7$^\circ$, 5.5$^\circ$, and
$5.5^\circ$ (Fishman et al. 1993).  The first value might seem
surprising, but $\theta_{\rm stat}$ is inversely proportional to the
square root of the burst fluence, not the burst brightness (flux).
Many faint Type I bursts have long durations and therefore small
$\theta_{\rm stat}$.  The typical total error $\theta_{\rm err} =
(\theta_{\rm sys}^2 + \theta_{\rm stat}^2)^{1/2}$ in the locations of
the bursts is thus 6.8$^\circ$.

One can show that $N_r$ recurrences of an individual source produce a
peak in the differential nearest neighbour distribution at $\theta_r =
\theta_{\rm err}/\sqrt{1.14 N_r} \approx 6^\circ/\sqrt{N_r}$, where we
have taken $\theta_{\rm err} = 6.8^\circ$.  The burst samples we study
all cluster on an angular scale $\approx 5^\circ$.  This implies $N_r
\simgreat 2$, and suggests that many of the bursts contributing to the
excess in the nearest neighbour distribution at small angular
separations are multiple recurrences of individual sources.

In a separate paper (Quashnock and Lamb 1993), we show that the 55 type
I bursts in the brightness range $465 < B < 1169$ exhibit a Galactic
dipole moment and a deviation of the Galactic quadrupole moment from
1/3 whose joint significance is high ($Q$-value = $1.1 \times 10^{-4}$,
taking into account having performed running averages); these
``medium'' brightness type I bursts are strongly concentrated toward
the Galactic center and in the Galactic plane.  They differ in this
respect from the bright ($B > 1900$ counts) type I bursts and the other
faint ($B \le 1900$ counts) type I bursts, which evidence hints lie
preferentially toward the Galactic anticenter but not in the Galactic
plane.  The distribution of type I $\gamma$-ray bursts on large angular
scales is thus a complicated function of burst brightness $B$.

In order to investigate further the relationship between the medium
type I bursts and other bursts, we plot the nearest neighbour
distributions {\it between} these medium type I bursts and other
brightness sub-samples.  We also plot the nearest neighbour
distributions between bright type I bursts and other brightness
sub-samples, because the statistical error is small compared to the
systematic error in the locations of the bright type I bursts, and the
uncertainty in their locations is therefore only about $4^\circ$.
Fig.~3 shows the resulting distributions and compares them with that
expected for a random distribution having the same number of bursts.
Table 3 gives the significances of the deviations of each from that
expected for a random distribution.

Fig.~3 (top panel) shows that bright type I bursts have no more medium
type I nearest neighbours at any separation than expected for a random
distribution (Q-value = 0.71).  Fig.~3 (bottom panel) shows that medium
type I bursts have no more nearest neighbours of all other types at any
separation than expected for a random distribution (Q-value = 0.66).
In contrast, bright type I bursts have significantly more faint type I
and II nearest neighbours within $\approx 5^\circ$ than expected for a
random distribution (Q-value = $4.0 \times 10^{-3}$).  We find that
this is also the case for the sample of faint type I and II bursts,
less the medium type I bursts (Q-value = $4.2 \times 10^{-3}$).  Thus
faint bursts cluster near themselves and near bright type I bursts, but
medium type I bursts show no significant correlation with bright type I
bursts or with other faint bursts.

We note that previous experiments were sensitive only to bright bursts,
and therefore did not detect the faint type I and type II bursts from
which the evidence comes that $\gamma$-ray burst sources repeat.

\section{Discussion}

We find that while the 260 bursts in the publicly available BATSE
catalogue are only modestly clustered, the 202 bursts in this sample
for which the statistical error in their location is $<9^\circ$ are
significantly clustered on an angular scale $\approx 5^\circ$.  We find
that the 201 bursts for which $B$ and $V$ exist and the 159 bursts in
this sample for which the statistical error in their location is
$<9^\circ$ are also significantly clustered on an angular scale
$\approx 5^\circ$.  This angular scale is smaller than the typical
(statistical plus systematic) error in burst locations of $6.8^\circ$,
suggesting multiple recurrences from individual sources.  We conclude
that ``classical'' $\gamma$-ray burst sources repeat on a time scale of
months, and that many faint type I and II bursts come from the sources
of bright type I bursts.  The values $D = 0.1877$ from the nearest
neighbour analysis within all bursts and $D = 0.2862$ from the analysis
between bright type I bursts and faint type I and II bursts imply that
at least $(260/202) (D/2) \approx$ 12\% (about 30) of all bursts and $D
\approx$ 30\% (about 12) of bright type I bursts detected during the 10
months of BATSE observations come from sources that repeated.

Many Galactic models (such as episodic accretion, starquakes, or
thermonuclear flashes involving neutron stars) predict repeated bursts,
whereas most cosmological models invoke a singular, cataclysmic event
(such as coalescence of neutron star-neutron star or neutron star-black
hole binaries) in order to generate the tremendous amount of energy
these models require.  The differences between the durations, spectra,
and time histories of many of the clustered bursts rules out the
possibility that gravitational lensing causes the repetitions.  Thus
the repeating nature of $\gamma$-ray burst sources favors Galactic
models, in agreement with the conclusion from our study of the Galactic
dipole and quadrupole moments of type I bursts that $\gamma$-ray bursts
are Galactic in origin (Quashnock and Lamb 1993).

The discovery that ``classical'' $\gamma$-ray bursts are Galactic in
origin (Quashnock and Lamb 1993) and recur, and the recognition that
many of them (i.e., the type II bursts) are of short duration
(Klebesadel 1992, Kouveliotou et al. 1993, Lamb and Graziani 1993a),
suggests that the distinctions between the sources of ``classical''
$\gamma$-ray bursts and ``soft $\gamma$-ray repeaters'' are dissolving;
only an apparent difference in spectral hardness remains, if that (see,
e.g., Fenimore et al. 1992).

Faint type I bursts are as much as $\sim$ 100 times fainter than bright
type I bursts, implying that the multiple bursts from individual
sources have a broad luminosity function.  Hence the limits on
recurrence time scales derived earlier (Schaefer and Cline 1985,
Atteia et al. 1987) are consistent with the repeating behavior we
find.  Further, the angular and brightness distributions of faint
bursts are strongly affected by the repeating nature of burst sources,
i.e., they largely reflect the burst luminosity function rather than
the spatial distribution of the sources.  Nevertheless, the significant
variations in the Galactic dipole and quadrupole moments of the type I
bursts as a function of $B$ (Quashnock and Lamb 1993) imply that the
(peak) brightness of many bright type I bursts is a surprisingly good
``standard candle'' and provides a good estimate of the distance to the
burst source.  Otherwise the variations in the angular distribution
would be washed out by the scatter in the intrinsic luminosity of the
bursts.

Elsewhere we conjecture that BATSE is seeing type I burst sources out
to $\sim 1 - 2$ kpc, and that the variations in the angular
distribution of these bursts as a function of burst brightness $B$
reflect the structure of the nearby spiral arms of the Galaxy
(Quashnock and Lamb 1993).  In this picture, the bright type I bursts
come primarily from the vicinity of the Orion arm, which lies around us
and toward the Galactic anticenter (at distances of up to $\approx$ 1
kpc).  The medium type I bursts come primarily from the vicinity of the
Sagittarius arm, which lies toward the Galactic center (at a distance
$\approx$ 1 - 1.5 kpc).  While some of the faint type I bursts may come
from the vicinity of the Perseus arm, which lies toward the Galactic
anticenter (at distances $\simgreat$ 2 kpc), most come from nearby
sources in the Orion spiral arm.

The findings of the present paper that medium type I bursts show no
significant correlation with bright type I bursts or with other faint
bursts, and that many faint type I and II bursts come from the sources of
bright type I bursts, support this picture.  In particular, they show
that the sources of the medium type I bursts are mostly distinct in
location from the sources of the other bursts, and that the angular
isotropy of the faint bursts reflects the fact that most come from the
(nearby) sources of bright type I bursts.

After this work was largely complete, we learned that Wang and
Lingenfelter (1993) find evidence that five particular bursts in the
BATSE catalogue are clustered on the sky and in time.  They suggest
that these bursts arise from a single repeating source.

\subsection*{Acknowledgments}

We gratefully acknowledge the contributions of the scientists who
designed, built, and flew BATSE on the {\it Compton} Observatory, and
whose efforts made possible the work reported here.  We thank Virginia
Wang and Rich Lingenfelter for making available to us the results of
their study in advance of publication.  We thank Carlo Graziani for
informative discussions about the nature of the clustering produced by
repeating sources.  Finally, we thank the anonymous referee for
pointing out that the Kolmogorov-Smirnov test is not valid for nearest
neighbour separations {\it within} samples.  This research was
supported in part by NASA grants NAGW-830, NAGW-1284, and NASW-4690.

\vfill\eject

\vfill\eject

\begin{figure}
\centering
\caption{
Distribution on the sky of the 260 bursts in the publicly available
BATSE catalogue (Fishman \etal\ 1993), in Galactic coordinates (the
Galactic Center lies at the center of the map).
}
\vskip 3cm
\end{figure}

\begin{figure}
\centering
\caption{
Comparison of the observed cumulative nearest neighbour distribution
for various burst samples and the expected distribution for a random
distribution with the same number of bursts.  The quantity $y \equiv 1
- \cos\theta$.  (top panel) All bursts in the publicly available BATSE
catalogue (heavy histogram) and all bursts for which the statistical
error in their locations is $< 9^\circ$ (light histogram) .  (bottom
panel) Faint type I bursts, all type I bursts, and all bursts for which
$B$ and $V$ exist, respectively.
}
\end{figure}

\begin{figure}
\centering
\caption{
Comparison of the observed cumulative nearest neighbour distribution
between various burst samples and the expected distribution for a
random distribution with the same number of bursts.  The quantity $y
\equiv 1 - \cos\theta$.  (top panel) Bright type I versus medium type I
bursts.  (bottom panel) Medium type I bursts versus other type I and II
bursts, bright type I versus faint type I and II bursts, and bright
type I versus faint type I and II less medium type I, respectively.
}
\vskip 9cm
\end{figure}

\null
\vfill\eject
\onecolumn

\begin{table}
\centering
\caption{Results of the Kolmogorov--Smirnov test for the cumulative
nearest--neighbour distribution within various samples of bursts.}
\medskip
\begin{tabular}{lcccccccc}
\hline
\hline
Sample & Number & D & $\theta_D(^\circ)$ & Q-value & Number & D &
$\theta_D(^\circ)$ & Q-value\\
\hline
\\
Faint Type I & 125 & 0.1714 & 4.4 & $1.3\times 10^{-2}$ &&&&\\
All Type I & 163 & 0.1642 & 4.8 & $4.8\times 10^{-3}$ &&&&\\
All Type I \& II & 201 & 0.1584 & 4.8 & $2.1\times 10^{-3}$ & 159 & 0.1903 &
4.8 & $7.9\times 10^{-4}$\\
All & 260 & 0.1168 & 4.9 & $1.8\times 10^{-2}$ & 202 & 0.1877 & 4.5 &
$1.1\times 10^{-4}$\\
\hline
\end{tabular}

Faint denotes bursts with $B <$ 1900 counts.  The right hand columns
are for the bursts in each sample for which the statistical error in
their location is less than 9$^\circ$. $D$ is the maximum deviation of
the cumulative nearest neighbour distribution from that expected for a
random distribution with the same number of bursts, and $\theta_D$ is
the angular separation at which it occurs.
\vskip 3cm
\end{table}

\begin{table}
\centering
\caption{Results of the Kolmogorov--Smirnov test for the cumulative
nearest--neighbour distribution between various samples of bursts.
The Medium type~I sample contains 55 type I bursts with $465<B<1169$ (counts)
found to have highly significant galactic dipole and quadrupole moments
(Quashnock and Lamb 1993).}
\medskip
\begin{tabular}{lclcccc}
\hline
\hline
Sample 1 & Number & Sample 2 & Number & D & $\theta_D(^\circ)$ & Q-value\\
\hline
\\
Bright Type I & 38 & Medium Type I & 55 & 0.1136 & 3.3 & 0.71\\
Medium Type I & 55 & Other Type I \& II & 146 & 0.0983 & 7.9 & 0.66\\
Bright Type I & 38 & Faint Type I \& II & 162 & 0.2862 & 5.0 & $4.0\times
10^{-3}$\\
Bright Type I & 38 & Faint Type I \& II & 107 & 0.2849 & 5.1& $4.2\times
10^{-3}$\\
&&less Medium Type I&&&\\
\hline
\end{tabular}

Faint, bright, and medium denote bursts with $B <$ 1900 counts, $B >$
1900 counts, and 465 counts $< B <$ 1169 counts.  $D$ is the maximum
deviation of the cumulative nearest neighbour distribution from that
expected for a random distribution with the same number of bursts, and
$\theta_D$ is the angular separation at which it occurs.
\end{table}
\end{document}